# Single photon Michelson-Morley experiment via de Broglie-Bohm picture: An interpretation based on the hypothesis of frame dragging

Masanori Sato

*Honda Electronics Co., Ltd.,*
*20 Oyamazuka, Oiwa-cho, Toyohashi, Aichi 441-3193, Japan*

E-mail: msato@honda-el.co.jp

**Abstract:** The Michelson-Morley experiment is considered via a single photon interferometer and a hypothesis of the dragging of the permittivity of free space $\varepsilon_0$ and permeability of free space $\mu_0$. The Michelson-Morley experimental results can be interpreted using de Broglie-Bohm picture. In the global positioning system (GPS) experiment, isotropic constancy of the speed of light, $c$, was confirmed by direct one way measurement. That is, Michelson-Morley experiments without interference are confirmed every day; therefore the hypothesis of frame dragging is a suitable explanation of the Michelson-Morley experimental results.

1. Introduction

In this discussion, the speed of light $c$, mass m, and length $x$ are assumed to be invariant. The physical reality is only time dilation by velocity. Mass and length look to be variant through the Lorentz transformation of reference time. Table 1 shows the comparison between orthodox interpretation and this proposal. Invariant and variant indicate the dependence on the velocity. The critical difference is that Lorentz contraction is not assumed, however the absolute stationary state is assumed to be indispensable. This discussion will be carried out within the theory of special and general relativity. Terms 1 and 2 in Table 1 are assumed. The assumption of the absolute stationary state and absolute velocity is independent on the theory of special relativity. That is, the absolute stationary state and absolute velocity are useful to eliminate paradoxes in the discussion of the theory of special relativity.

Table 1 Comparison between orthodox interpretation and this proposal

|   | Terms | Orthodox interpretation | This proposal |
|---|---|---|---|
| 1 | The speed of light: $c$ | Invariant | Invariant |
| 2 | Reference time: $t$ | Variant | Variant |
| 3 | Mass: m | Variant and Invariant | Invariant |
| 4 | Velocity: $u$ | Relative between two observers | Variant: not relative between two observers |
| 5 | Lorentz contraction | Yes | No |
| 6 | Absolute stationary state | No | Yes |
| 7 | Absolute velocity | No | Yes |

Let us consider the velocity *u*, for which, in this discussion the stationary state and moving frame are defined. These experiments are carried out in the gravitational field of the earth. **Figure 1** shows the following illustration: the stationary state is at the north pole of the earth and the moving frame is the global positioning system (GPS) satellite. For a more simplistic example, place observer A is located at the North Pole in order to eliminate the earth's rotation, and neglect the effect of the gravitational potential of the earth. The gravitational field of the earth can be **provisionally** considered as the stationary state. A velocity of 30 km/s in the solar system does not affect the GPS satellite or the observer on earth. Thus, the velocity of the GPS satellite $u_G$ =4 km/s observed in the earth-centered locally inertial (ECI) coordinate system can be assumed to be the absolute velocity. Observer A on earth sees a time dilation on the clock in the GPS satellite; however, the observer in the GPS satellite sees not a time dilation but a time gain on the clock on earth. This is because only the reference time in the GPS satellite becomes large.

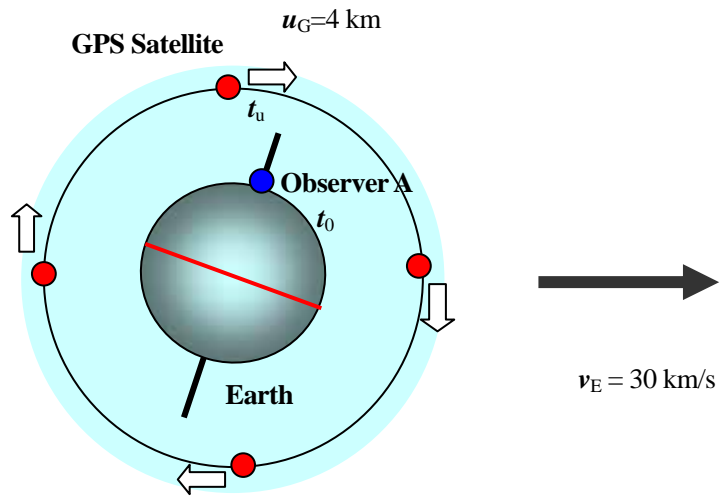

**Fig. 1** Illustration of the stationary and moving frames: The stationary state is on the north pole of the earth and the moving frame is the GPS satellite. The velocity of the earth in the solar system $v_E$=30 km/s does not affect the earth or the GPS satellite. The velocity of the GPS satellite $u_G$ =4 km/s observed in the ECI coordinate system can be assumed to be the absolute velocity. The time dilation of the GPS satellite is 7.1 μs every day

The reference time expansion according to the Lorentz transformation indicates that the phenomena progress slowly when seen from a stationary state. In **Fig. 1**, the time dilation by velocity only occurs on the GPS satellite. Observer A does not suffer the time dilation by the velocity. This is because observer A is **provisionally** in a stationary state (the effect of the gravitational potential is neglected.). The GPS satellite orbits the earth, and thus, only a transverse Doppler shift is detected. The difference of the reference time is calculated as follows,

$$\frac{t_0 - t_u}{t_0} = 1 - \frac{1}{\sqrt{1 - \left(\frac{4}{300000}\right)^2}} = -0.889 \times 10^{-10}.$$

Time dilation of the GPS satellite is $0.889 \times 10^{-10} \times 60 \times 60 \times 24 = 7.1 \mu s$ every day [8]. Therefore,

$u_G$ =4 km/s is not relative. That is the GPS satellite detects a faster velocity of observer A. This is because the reference time of the GPS satellite is expanded, thus the GPS satellite observes the lager velocity of observer A.

In 1881, from the first interferometer experimental results, Michelson [1] described that "the result of the hypothesis of a stationary ether is thus shown to be incorrect." The hypothesis that the earth dragged the ether along with it in its orbit was proposed, accounting for the negative result of the interferometer experiment. In those days, Lord Rayleigh wrote to Michelson to check the hypothesis of "aether drag" proposed by Fresnel. Michelson only denied the stationary ether. However, there was no direct experimental evidence for the existence of the ether, thus it has been widely considered that there is no medium in space. That is, the theory of special relativity can explain everything without the ether. Thus the hypothesis of ether dragging seemed to disappear. However, there is no experimental data to deny the hypothesis of ether dragging.

The hypothesis of frame dragging appeared again in 1918. The Lense-Thirring effect [2], that is the rotational frame dragging, was derived from the theory of general relativity. Although accelerated or rotational frame dragging effects were proposed, however there were few discussions of the frame dragging in the inertial motion of the gravitational field. It is described that inertial frame dragging is usually associated with a gravitational field generated by rotating matter [3]. The hypothesis of ether dragging, which was caused by the Michelson paper in the 1880s, looks very interesting, and we should reconsider the remarkable hypothesis of ether dragging.

Quantum mechanics and special relativity have a very interesting relationship, although the concepts of these two theories appear to be quite different. This is because quantum mechanics and special relativity can be discussed via the phenomenon of interference. We consider that quantum mechanics and special relativity both have, as a basis for their theories, experimental data on interference. This is the reason why they have a particular compatibility. The relationship between energy E and momentum M as E=M/c (c: the speed of light) is one suitable examples.

The Michelson-Morley experiment demonstrates that an interference pattern does not vary according to the motion of the earth. This is the fundamental data of the theory of special relativity. Thus the phenomenon of interference is critically related to the concept of special relativity, and is also related to quantum mechanics. I consider that the phenomenon of interference is the key to establishing a conceptual relationship between quantum mechanics and special relativity.

I have previously pointed out that another means of interpretation is possible for the Michelson-Morley experiment from the view point of the de Broglie-Bohm picture [4]. I have also discussed time dilation in an atomic clock in motion [5]. It is important to investigate the Michelson-Morley experiment with respect to new understandings, for example the GPS experiment.

In these days, the accuracies of the experiments based on the GPS [6] have been incredible, and therefore, we should reexamine the historic hypotheses. The GPS experiment is considered to be carried out in the special condition of the gravitational field of the earth. In the GPS experiment, the isotropic constancy of the speed of light is confirmed but the evidence of the Lorentz contraction is not observed [7]. That is, the earth is considered as if in the absolute stationary state. Therefore the null result of the Michelson-Morley experiment is reasonably explained. The ECI coordinate system seems to be the absolute stationary coordinate system that supports the GPS experiments [8].

In this letter, we note that the Michelson-Morley experiment shows the interference of photons;

however, it does not show the photons' simultaneous arrival at the beam splitter (half mirror). The experiment also revealed that the interference conditions were not affected by the motion of the earth. Thereafter we show a hypothesis of the dragging of the permittivity of free space $\varepsilon_0$ and permeability of free space $\mu_0$ [9].

2. Single photon Michelson interferometer: interpretation via de Broglie-Bohm picture

**Figure 2** shows the schematic diagram of a single photon Michelson interferometer. A photon enters the interferometer via the beam splitter, is reflected by the mirror, and is then recombined by the beam splitter. We can detect the interference, that is, the photon paths can be arranged such that the detector detects the photons.

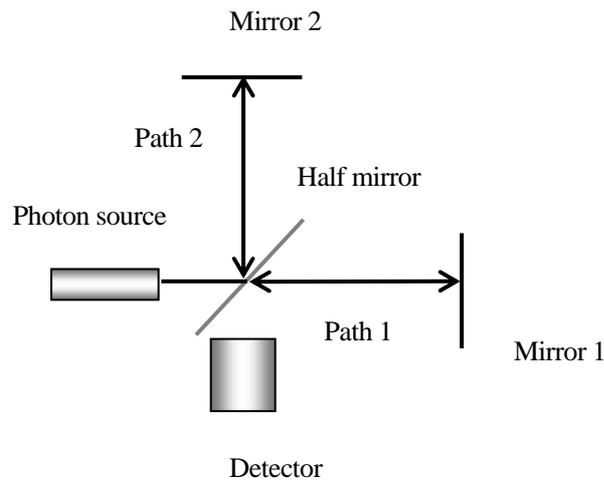

**Fig. 2**    Conceptual diagram of Michelson-Morley experiment

**Figure 3** shows a schematic diagram of the quantum potential. According to this schematic diagram, a single photon Michelson interferometer appears to detect only the interference, and does not measure the speed of photons. In a single photon interferometer, there is only a photon in the photon paths, therefore it cannot measure the arrival time of photons on path 1 and path 2. Only the interference condition is evaluated.

The de Broglie-Bohm picture [10, 11] provides a suitable means of interpreting this situation. The phenomenon is explained using the wave and particle model. The wave is the pilot wave (quantum potential) and the particle is a photon that is guided by the pilot wave [10, 11]. The wave is nonlocal and the particle, that is, the photon, is local. The photon travels at the speed light, namely, the photon has compatibility with special relativity; however, the wave does not have compatibility with special relativity (i.e., nonlocal). Interference is determined by the geometry of the interferometer, as shown in **Fig. 3**, i.e., the interference is governed by the lengths of paths 1 and path 2. It should be pointed out that the Michelson-Morley experiment does not measure the velocity of the photon.

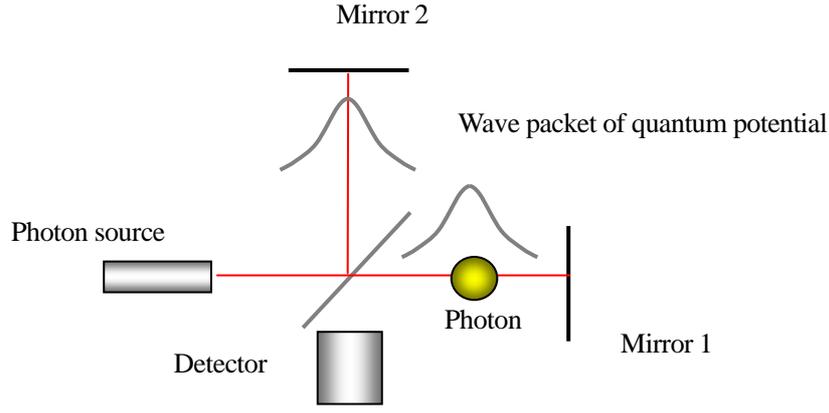

**Fig. 3** Interpretation using de Broglie-Bohm picture. Using single photon interferometer, Michelson-Morley experiment can be carried out. Interpretation using the de Broglie-Bohm picture, that is, two wave packets of quantum potential define the interference.

3. Derivation of the Lorentz contraction

The derivation of the Lorentz contraction is described as follows [12]. **Figure 4** shows the Michelson-Morley experimental setup in free space. According to the red arrows,

$$(ct_1)^2 = L^2 + (vt_1)^2. \tag{1}$$

$$t_1 = \frac{L}{\sqrt{c^2 - v^2}}. \tag{2}$$

From the blue arrows,

$$t_2 + t_3 = \frac{L_L}{c-v} + \frac{L_L}{c+v} = \frac{2L_L/c}{1 - \left(\frac{v}{c}\right)^2}. \tag{3}$$

Where, $L_L$ shows the contracted length by Lorentz contraction. Equations (1) to (3) are valid in Minkowski space (free space). The Michelson-Morley experiment in the gravitational field of the earth showed simultaneous arrival of two photons represented as follows,

$$2t_1 = t_2 + t_3. \tag{4}$$

If we apply equations (2) and (3) to equation (4), the Lorentz contraction is derived as,

$$L_L = \sqrt{1 - \left(\frac{v}{c}\right)^2} L. \tag{5}$$

Equation (5) shows the Lorentz contraction. I do not agree with the derivation of equations (3) and (4), or, therefore, (5). Equation (3) is not compatible with the experimental results of the GPS experiment. Equation (4) cannot be derived from the Michelson-Morley experiment, where there was no fringe shift of interference pattern. Derivation of equation (5) will be discussed in section 4.

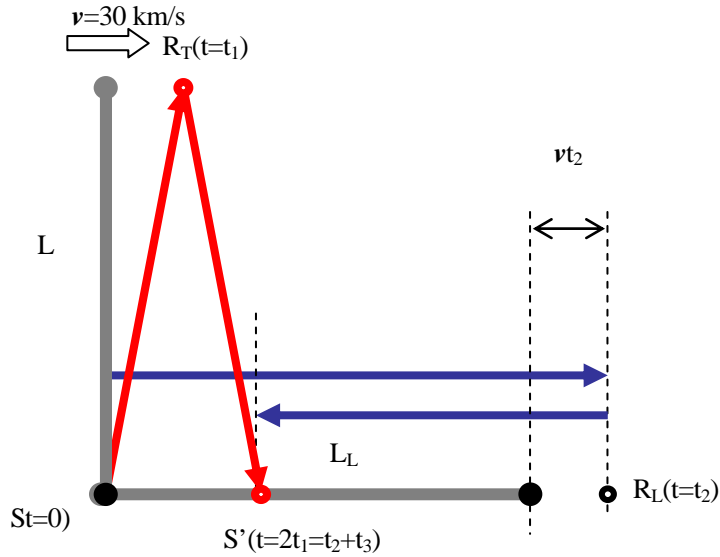

Fig. 4 Derivation of the Lorentz contraction from the Michelson-Morley experimental setup. Length parallel to v is assumed to contract as $L_L$.
S: photon source (at t=0)
S': photon source (at $t=2t_1=t_2+t_3$)
L: path length
$R_T$: reflector of photon transverse to v (at $t=t_1$)
$R_L$: reflector of photon parallel to v (at $t=t_2$)
$t_1$: flight time of photon from S to $R_T$
$t_2$: flight time of photon from S to $R_L$
$t_3$: flight time of photon from $R_L$ to S'

4. Reconsideration of the derivation of the Lorentz contraction from the GPS experiment

The ECI coordinate system works as if the earth is substantially in an absolute stationary state. This is derived from the experimental data of the atomic clock in the GPS satellite and isotropic constancy of the speed of light. This is a reconsideration of the hypothesis of frame dragging by the gravitational field which satisfies the assumption that "the speed of light, *c* is constant regardless of the velocity of the light source and the observer".

 If this hypothesis is used, the Michelson-Morley experiment is easily explained because the gravitational field of the earth is considered as the absolute stationary state. It was discussed that the Michelson interferometer cannot detect any earth motion because the interference does not show the simultaneous arrival of two photons. This is because we can obtain the same result using a single photon interferometer. Single-photon experiments do not show the simultaneous arrival of two photons.

However, the Michelson-Morley experimental results were reconfirmed in the GPS experiment. Isotropic constancy of the speed of light was confirmed not by interference experiments but by one way direct measurements as shown in **Fig. 5**. In the GPS experiment, measurement is done by single path without interference.

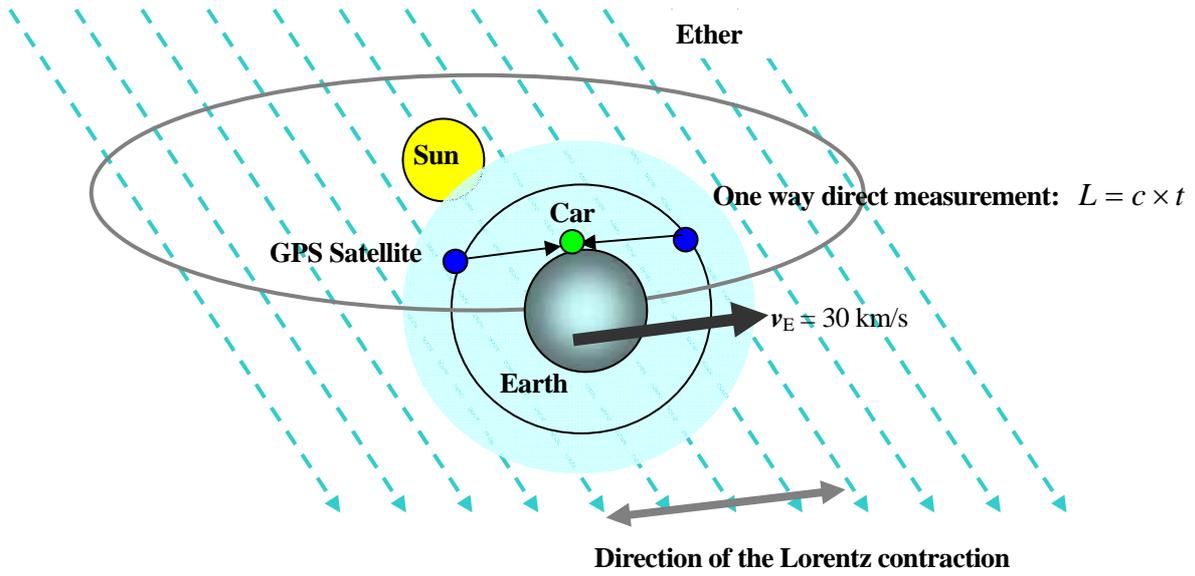

Fig. 5 Illustration of the earth motion in the solar system and frame dragging. The earth motion is assumed to be 30 km/s. At a moment when a car on earth faces two GPS satellites, the car measures the distance from two GPS satellites. One way direct measurement from the GPS satellite to the car, that is, $L = c \times t$, is applicable. The velocity $v_E$ = 30 km/s in the solar system does not affect on the measurement.

In the derivation of the Lorentz contraction equation (3) is assumed. The values in equation (3) are measured respectively, as represented in equation (6),

$$\frac{L}{c - v_E}, \quad \frac{L}{c + v_E}. \tag{6}$$

Where, $v_E$ is the velocity of the earth in the solar system. These values in equation (6) were assumed in the derivation of the Lorentz contraction. However, they were not observed in the GPS experiment, which was described as follows,

$$L = ct. \tag{7}$$

Equation (7) shows that we only detect the speed of light c. The effects of the velocity $v_E$ were not observed. Therefore, equation (3) cannot be used. The sensitivity of a single path measurement in the GPS experiment is $30 km \div 300,000 km = 10^{-4}$, thus, is higher than that of the Lorentz contraction calculated as $1 - \dfrac{1}{\sqrt{1 - \left(\dfrac{30}{300000}\right)^2}} = -0.500 \times 10^{-8}$. Isotropic constancy of the speed of light in the GPS experiment was confirmed experimentally. Thus, the isotropic constancy of the speed of light was observed not only by interference experiment, but also by direct measurement in GPS experiments.

The GPS experimental results are summarized as follows: the arrival time of photons are equal, and thus are represented as,

$$t_1 = t_2 = t_3. \tag{8}$$

Equation (8) satisfies equation (4). Using the principle of isotropic constancy of the speed of light, the path lengths of the red arrows and two blue arrows are completely equal. Thus the illustration in **Fig. 4** cannot be applied to the Michelson-Morley experiment. This, illustration can be applicable only to Minkowski space (free space). **Figure 6** is the illustration of the Michelson-Morley experiment in the gravitational field of the earth. This illustration is that of in the absolute stationary state.

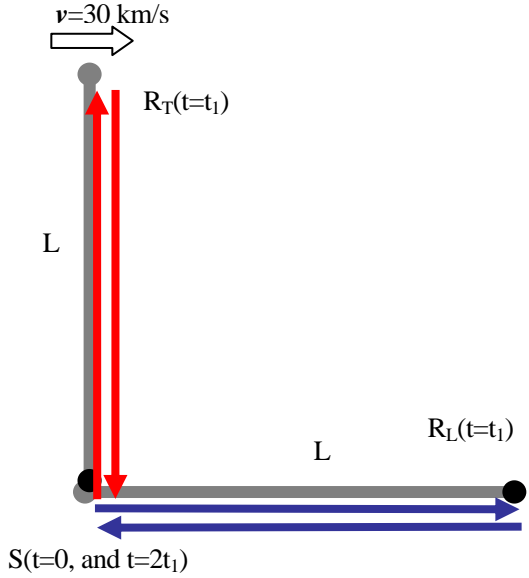

**Fig. 6** Michelson-Morley experimental setup in the gravitational field of the earth. Frame dragging hypothesis can explain not only the Michelson-Morley experiment but also the GPS experiment. The single photon Michelson-Morley experiment can also be explained.

Therefore, the assumption of the Lorentz contraction does not reasonably explain the Michelson-Morley experiment. One of the alternative hypotheses is frame dragging.

5. Hypothesis of frame dragging

Now let us discuss frame dragging. The GPS experiments as well as the Michelson-Morley experiments were carried out in the gravitational field of the earth. A hint for frame dragging is obtained from the fact that acoustic waves in the atmosphere are not affected by the motion of the earth. The phase velocities of an electromagnetic wave and acoustic wave are summarized in Table 2. One of the reasons why the ether was denied comes from the analogy of the acoustic wave: that is, if *c* is very large, then a large stiffness C is required. It is rather difficult to assume the drag of a large stiffness. However, from the equation of the velocity of electromagnetic waves frame-dragging indicates that the permittivity of free space $\varepsilon_0$ and the permeability of free space $\mu_0$ are dragged. At this stage, I do not have an idea for the frame-dragging, but nevertheless this hypothesis seems to be reasonable.

If the frame dragging hypothesis is adopted, the Lorentz contraction is not derived. The Michelson-Morley experimental results and GPS experimental results are simultaneously explained. The single photon Michelson-Morley experiment is also explained. This is because interference critically depends on the path length. If there is no shift of interference pattern (fringe shift), the path lengths are equal, as shown in **Fig. 6**.

Table 2 Comparison of electromagnetic and acoustic waves

|   | Waves | Phase velocity | Coefficients |
|---|---|---|---|
| 1 | Electromagnetic wave | $c = \dfrac{1}{\sqrt{\varepsilon_0 \mu_0}}$ | $\varepsilon_0$: permittivity of free space <br> $\mu_0$: permeability of free space |
| 2 | Acoustic wave | $c_A = \sqrt{\dfrac{C}{\rho}}$ | C: coefficient of stiffness <br> $\rho$: density |

6. Conclusion

We discussed the Michelson-Morley experiment that employs a single photon interferometer, and pointed out that this experiment demonstrates that interference is independent of the motion of the earth. We also proposed a means of interpretation of this experiment using de Broglie-Bohm picture. Furthermore, to explain the GPS experimental results of the isotropic constancy of the speed of light, the hypothesis of frame dragging is reconsidered. Frame dragging indicates the dragging of the permittivity of free space $\varepsilon_0$ and permeability of free space $\mu_0$. The hypothesis of frame dragging seems reasonable, and the ECI coordinate system is substantially in the absolute stationary state.

References
1) A. Michelson, Am. J. Sci, **122,** 120 (1881).
2) J. Lense, and H. Thirring, Physikalische Zeitschrift, **19**, 156-63, (1918).
3) R. Mallette, "Weak gravitational field of the electromagnetic radiation in a ring laser," Phys. Lett, A, **269**, 214, (2000).
4) M. Sato, "Proposal of Michelson-Morley experiment via single photon interferometer: Interpretation of Michelson-Morley experimental results using de Broglie-Bohm picture," physics/0411217, (2004).
5) M. Sato, "Proposal of atomic clock in motion: Time in moving clock, " physics/0411202, (2004).
6) N. Ashby, "Relativity in the Global Positioning System," www.livingreviews.org/Articles/Volume6/2003-1ashby, (2003).
7) M. Sato, "Incompatibility between the principle of the constancy of the speed of light and the Lorentz contraction in the GPS Experiment," physics/0703123, (2007).
8) M. Sato, "Interpretation of special relativity as applied to earth-centered locally inertial coordinate systems in Global Positioning System satellite experiments," physics/0502007, (2005).
9) M. Sato, "Doppler shift between two moving gravitational fields: A hypothesis of the dragging of the permittivity $\varepsilon_0$ and permeability $\mu_0$," arXiv:0704.1942, (2007).
10) D. Bohm and B. Hiley, *The Undivided Universe,* (Routledge, London, 1993).
11) P. Holland, *The Quantum Theory of Motion,* (Cambridge University Press, Cambridge, 1994).
12) R. Feynman, R. Leighton, and M. Sands, "*The Feynman Lectures on Physics*," (Addison Wesley, Reading, MA, 1965), Vol. 2.